\documentclass[9pt,a4paperpaper,runningheads]{article}

\usepackage{lmodern}
\usepackage{amssymb,amsmath}
\usepackage{ifxetex,ifluatex}
\IfFileExists{microtype.sty}{%
\usepackage[]{microtype}
\UseMicrotypeSet[protrusion]{basicmath} 
}{}

\usepackage{hyperref}
\usepackage{graphicx,grffile}
\usepackage{xcolor}

\usepackage{graphicx,grffile}
\makeatletter
\def\maxwidth{\ifdim\Gin@nat@width>\linewidth\linewidth\else\Gin@nat@width\fi}
\def\maxheight{\ifdim\Gin@nat@height>\textheight\textheight\else\Gin@nat@height\fi}
\makeatother
\setkeys{Gin}{width=\maxwidth,height=\maxheight,keepaspectratio}
\makeatletter
\def\fps@figure{htbp}
\makeatother
\providecommand{\tightlist}{%
  \setlength{\itemsep}{8pt}\setlength{\parskip}{0pt}}
\usepackage{authblk}
\usepackage[a4paper,margin=1.2in]{geometry}
\providecommand{\keywords}[1]{\textbf{\textit{Keywords:}} #1}

\usepackage[square,numbers,sectionbib]{natbib}
\ifluatex
\else
  \SetExpansion
  [ context = sloppy,
    stretch = 30,
    shrink = 60,
    step = 5 ]
  { encoding = {OT1,T1,TS1} }
  { }
\fi
\usepackage{url}
\makeatletter
\g@addto@macro{\UrlBreaks}{\UrlOrds}
\makeatother

\title{DINGO: an ontology for projects and grants linked data}
\author[1]{Diego Chialva\footnote{Corresponding author}} 
\author[1]{Alexis-Michel Mugabushaka} 

\affil[1]{ERCEA\footnote{\textbf{Disclaimer.} The views expressed in this paper are the authors'.
They do not necessarily reflect the views or official positions of the
European Commission, the ERC Executive Agency or the ERC Scientific
Council.}, Place Charles Rogier 16, 1210 Brussels, Belgium\footnote{\textbf{Emails:} Diego-Valerio.Chialva@ec.europa.eu,
Alexis-Michel.Mugabushaka@ec.europa.eu}}
          
\date{}

\begin{document}
\maketitle
\begin{abstract}
We present DINGO (Data INtegration for Grants Ontology), an ontology
that provides a machine readable extensible framework to model data for semantically-enabled applications
relative to projects, funding, actors, and, notably, funding policies in
the research landscape. DINGO is designed to yield high modeling power
and elasticity to cope with the huge variety in funding, research and
policy practices, which makes it applicable also to other areas besides
research where funding is an important aspect. We discuss its main
features, the principles followed for its development, its community
uptake, its maintenance and evolution.
\keywords{ontology \and linked data, research funding, research projects, research policies}
\end{abstract}

\hypertarget{sec:introduction}{%
\section{Introduction, Motivation, Goals and
Idea}\label{sec:introduction}}

Services and resources built around Semantic Web, semantically-enabled applications and linked (open) data
technologies have been increasingly impacting research and
research-related activities in the last years. Development has been
intense along several directions, for instance in ``semantic
publishing'' \citep{S2009SP}, but also in the aspects directed toward
the reproducibility and attribution of research and scholarly outputs,
leading also to the interest in having Open Science Graphs
interconnected at the global level \citep{RDAOSGF2019}. All this has
become more and more essential to research practices, also in light of
the so-called reproducibility crisis affecting a number of research
fields (see, for instance, the huge list of latest studies at
\href{https://reproduciblescience.org/2019}{\emph{https://reproduciblescience.org/2019}}).

In fact, the demand of easily and automatically parsable, interoperable
and processable data goes beyond the purely academic sphere. The
research landscape comprises a vast number and type of activities, with
multiple and diverse stakeholders, actors and with impact on several
aspects and sectors of society. One aspect of huge relevance is the
funding of research, together with the related policies for science
development and sustainability.

Machine-actionable, inter-operable data is in huge demand in those
respects. On the one hand, for instance, research funding agencies face
increasing pressure to report on impact derived from their activities.
This has to be seen in a broader context of the increased role that
research assessment play in research policy debates. On the other hand,
researchers and research organisations are asked more and more to
conform to policy specifications in order to obtain and secure their
funding. The compliance to funding and research polices is also part of
the wider debate about best research practices such as Open Science,
Open Access, FAIR data and sustainable research.

Research assessment and compliance verification at any level involves
collection, management and analysis of a great increasing deal of data
of different types and from multiple sources . The classical way to meet
this demand has been to collect data directly from various research
actors. This increases the burden on researchers, university
administration and funding agencies, as those data has to be managed and
curated. Moreover, the information, typically collected in an ``ad hoc''
way and in isolation, is not available to others. This results also in
duplication of efforts, due to the necessity to re-do the linking and
processing of data. The difficulty of data linking and semantic
interpretation across different realities and agencies also entails that
data and analysis are of limited value when it comes to put them in
broader perspective.

Solving these problems entails having data that can be easily parsed,
processed and interpreted computationally. This requires expressive
shared machine-processable descriptions and models on the Web.
Technologies as RDF, RDFS, OWL, and SPARQL provide building blocks
towards that goal and have favoured the development of ontologies to
describe various aspects of the research domain.

However, the development of ontologies for the funding aspects of
research and their relations to research activities, actors is still
quite in its early stages. In particular, while few ontologies exist
(see section \ref{sec:relatedWork}), they mostly envisage only some of
the important semantic elements (typically those relative to projects
and grants), as we will show.

This note presents a novel ontology, developed to manage data on
research grants and projects, but also notably to conceptualise funding
policies and instruments, facilitating the integration and
interoperability of such information with other data and from various
sources in the framework of the so-called Linked Data. The ontology has
been dubbed DINGO (Data INtegration for Grants Ontology). It provides an
extensible, interoperable framework for formally modeling the relevant
parts of data in this knowledge area.

DINGO particularly facilitates the effort of putting analysis of funding
activities and policies in broader context and comparative perspectives,
which is much needed when assessing research, policies and their impact.
In this way, DINGO will be beneficial in practice at several levels. For
instance, by increasing the capacity of analysis to inform policy and
strategic discussions, as well as reducing the effort of researchers and
officers in giving evidence of policy compliance.

Indeed, one specific characteristic of the knowledge area DINGO aims to
describe is its variety. The existing funding activities and policies
show a large spectrum of practices, with remarkable diversity and
complex semantics. This constitutes a serious difficulty when trying to
put funding activities and policies in context and comparative
perspectives. DINGO has therefore been specially designed to cope with
this, by a rigorous conceptualisation of commonalities via a number of
ontology classes and properties, together with other classes that allow
tuning semantic specializations to the specific cases when modelling
data.

This also allows DINGO not only to be effectively used as a pure domain
ontology specific to research activities, but in fact to perfectly model
even other domains where funding activities play a relevant roles (such
as the arts, cultural conservation, and many others). DINGO has
therefore, in some respects, also the multi-domain usability typical of
more upper ontologies (we use here the classification ad definitions of
ontologies by Guarino \citep{G1997}).

DINGO is fully documented at
\href{https://w3id.org/dingo}{\emph{https://w3id.org/dingo}}, and a
machine readable version of the ontology is available at
\href{https:w3id.org/dingo}{\emph{https://w3id.org/dingo}} in RDF-Turtle
by redirection when visiting with the ``text/turtle'' header (it is also
available at
\href{https://dcodings.github.io/DINGO/DINGO-OWL.ttl}{\emph{https://dcodings.github.io/DINGO/DINGO-OWL.ttl}}).

This article is organised as follows. Section \ref{sec:relatedWork}
discusses related work. Sections \ref{sec:communityUptake},
\ref{sec:mappingReuseExtensions}, \ref{sec:dingoPresentation},
\ref{sec:maintenanceEvolution} and subsections thereof present the aims,
development guidelines, community uptake, maintenance and evolution, and
main features of DINGO (we leave the detailed description of the
ontology to its documentation, available online). We conclude in section
\ref{sec:conclusion}, where we also comment on future potential
directions of development.

\hypertarget{sec:relatedWork}{%
\section{Related Work}\label{sec:relatedWork}}

A few works exist modelling data related to funding and research,
although to our knowledge none has been dealing with the aspects
pertaining to research (funding) policies together with the rest.

One of the earliest efforts to create a data model for the management of
research funding data is CERIF (Common European Research Information
Format), \citep{CERIF2010}. It is an extremely rich and detailed
vocabulary for research management, with a considerable number of
entities and relations, and a high granularity. However, it does not
conceptualise aspects related to policies.

CERIF, conceived for CRIS (Current Research Information Systems), has
deep roots in relational database modeling more than in the
semantic/knowledge graph one, as visible from some of its
characteristics. For example, one of its main features is the presence
of ``link entities'' such as project-organisation, project-person, and
so on. They are in fact relationships rooted in relational database
reification practices (which differ from what reification is in the
framework of knowledge graphs and semantic web). Such ``link entities''
have however less straightforward interpretation in terms of semantic
concepts (they often represent couples of concepts), which would affect
inferences. We will show how DINGO avoids this problem and yet manages
to capture the aspects of interest.

Related to CERIF is the OpenAire data model \citep{OA2019DM}.
\emph{OpenAire} \citep{OpenAire} is an infrastructure that links
research outcomes to their creators, enabling discoverability,
transparency, reproducibility and quality-assurance. The OpenAire data
model uses part of the CERIF vocabulary (including some of the ``link
entities'') and combines them with the OpenAire guidelines.

A few OWL-based ontologies exist describing funding in research.
Compared to CERIF, they are fully framed in semantic modeling. The most
well-known ones (and in fact the only ones to our knowledge) are FRAPO
(Funding, Research Administration and Projects Ontology) \citep{FRAPO},
\citep{PSSD2018}, and the Springer Nature SciGraph Ontology
\citep{SCIGRAPH}.

These are actually part of larger ontologies or ontology collections
mainly aiming at categorizing scholarly data, such as publications and
other similar outputs, rather than focusing exclusively on the funding
and research landscape. They are thus tuned for those other purposes and
have specific limitations. For example, the SciGraph one does not appear
to distinguish the concept of ``grant'' as funding from the concept of
``research project'' and thus would not allow to easily model for many
existing funding practices and uses cases (for instance, the case of
projects with multiple grants, either co-occurring or in sequence).
FRAPO instead lacks classes and properties for relevant concepts such as
``principal investigators'' and others . Moreover, neither ontology
conceptualises the domain of funding policies.

In addition to these, there is a growing number of initiatives
addressing other dimensions of research data than the funding-project
ones. To cite a few: \emph{OpenCitations} \citep{OpenCitations}, which
is dedicated to open scholarship and open bibliographic and citation
data; \emph{SMS} (Semantically Mapping Science) \citep{SMS}, a platform
integrating heterogeneous datasets for science, technology, innovation
studies; \emph{VIVO} \citep{VIVO} an open source software and ontology
for representing scholarship and scholarly activity. Finally one can
mention also \emph{CASRAI} (Consortia Advancing Standards in Research
Administration Information) \citep{CASRAI}, which does not provide an
ontology, but a glossary of research administration information.

We will discuss the part of schema.org \citep{SCHEMAORG} dealing with
funding data in Section \ref{sec:communityUptake}, as it was in fact
inspired by DINGO.

We finally would like to mention the FP Ontologies \citep{FPOEGUPM}.
They do not deal with research funding, but model some aspects of
projects. Web-searching them points to the webpage at \citep{FPOEGUPM},
but in fact we could not find documentation nor download any
serialisation from that page.

\hypertarget{sec:communityUptake}{%
\section{Community Uptake and Use of DINGO}\label{sec:communityUptake}}

DINGO has been first presented to the public in the late 2018, and has
led to a number of uses, both directly for data modeling and knowledge
bases creation, and as a basis or inspiration for related ontology
modeling efforts.

The first public presentation of DINGO has occurred at the workshop
``Wikidata for research'', Berlin, 17-18 June 2018, where feedback and
input were exchanged with a working group of participants, which lead to
the linking of DINGO with the Wikidata graph.

DINGO also inspired the part of the schema.org model specific for grants
and funding (as mentioned explicitly at the issue 343 of the schema.org
release of 2019-04-01\footnote{Visible at
  \href{https://schema.org/docs/releases.html}{\emph{https://schema.org/docs/releases.html}}).}.
Schema.org 's model covers however only a subset of DINGO's .

Furthermore, DINGO has been adopted to model the knowledge base of the
European Commission data hosted and available now in the OpenAire LOD
service (at
\href{http://lod.openaire.eu/eu-open-research-data}{\emph{http://lod.openaire.eu/eu-open-research-data}}),
and as one of the basis of the schema for the GRANTID initiative of
CrossRef \citep{CROSSREF} (one of the authors of this article, D.C., has
been a member of the technical group for the schema\footnote{See
  \href{https://www.crossref.org/working-groups/funders/}{\emph{https://www.crossref.org/working-groups/funders/}}).}).

\hypertarget{sec:mappingReuseExtensions}{%
\section{Ontology Mapping, Reuse and Extensions in
DINGO}\label{sec:mappingReuseExtensions}}

Ontology mapping is a key challenge of the Sematic Web and of Linked
(Open) Data for several reasons. Ontology reuse is also a good knowledge
engineering practice, increasing the interoperability of systems.

In the framework of semantic modeling and the Semantic Web, reuse and
mapping are particularly complex. On the one hand, the de-centralised
nature of the web favours the development of several ontologies and data
models, which often overlap partially. On the other hand, the single
ontologies are generally created with specific goals, and thus even when
they are developed to model data from the same domain(s), they will
generally present subtle semantic differences even in seemingly general
concepts.

In the case of research data, mapping and reuse are further complicated
by the multiplicity of actors and the diversity of types of funding
practices, policies and data. But on the other hand, this same issue
prompts to maximise the semantic modeling power of an ontology by
linking it with overlapping ones in order to achieve maximum
interoperability.

DINGO was therefore built from the start with a particular attention to
ontology mapping. Pure reuse has been possible only to a certain extent,
because ontologies covering overlapping knowledge areas (such as those
mentioned in Section \ref{sec:relatedWork}) do indeed present subtle but
relevant semantic specificities.

The mapping in DINGO makes use of the SKOS ontology/data model mapping
properties (documented at \citep{SKOS}) and RDF and OWL class and
properties axioms such as owl:equivalentClass and owl:equivalentProperty
when applicable. In fact, the establishment of mapping using the latter
owl axioms is generally quite complicated, as they require establishing
that the full extension of the relative classes/properties are equal.
This is typically a difficult task in the case of a complex knowledge
area such as the one of research, and has therefore being done carefully
and rather conservatively in DINGO.

DINGO is presently mapped to the Wikidata data model, to schema.org and
to the FRAPO ontology. There is also interest in linking DINGO with the
vocabulary provided by CERIF, and future developments have been already
planned in that sense.

Besides that, DINGO also reuses several other ontologies, such as SKOS,
schema.org and DublinCore \citep{DCMI}, and is inspired by the FAIR
principles \citep{WDAAea2016} for data publication.

Finally, DINGO has been designed to be easily extensible to adapt to the
various possible use cases and diversity of data and existing practices.
The ontology presents ``hook properties'' (such as
\emph{product\_or\_material\_produced}) that allow to extend DINGO
linking, for instance, to data modeled with the many ontologies dealing
with scholarly and publishing data (such as the SPAR ontologies
\citep{PSSD2018}, the Semantic Web Journal (SWJ) ontology
\citep{HJMSH2013}, the Semantic Web Conference (SWC) ontology
\citep{SWC}, the Semantically Annotated LaTeX (SALT) ontologies
\citep{GHMD2007}, the Nature Ontologies \citep{HP2015}, the SciGraph
Ontologies \citep{HPT2017}, the Conference Ontology \citep{NGPG2016},
BIBFRAME \citep{BIBFRAME} and bibkliotek-o \citep{BIBLIOTEKO}).

\hypertarget{sec:dingoPresentation}{%
\section{The DINGO Ontology}\label{sec:dingoPresentation}}

\hypertarget{sec:ontologyAims}{%
\subsection{Aim of the Ontology}\label{sec:ontologyAims}}

The principal aim of the DINGO ontology is to provide a machine readable
extensible framework to model data relative to projects, funding,
policies and actors. The original intended users for such frameworks
were the stakeholders in the research landscape with their very
different use cases.

As discussed in Section \ref{sec:introduction}, semantical modeling of
that knowledge area faces, among others, one main difficulty: there
exist a huge variety of funding, policies, practices and research
activities. Due to the aim of being able to cope with this, as we
illustrate also in Section \ref{sec:ontologyDescription}, DINGO is
finally applicable also to domains different than the one of pure
research where the funding aspects are relevant, for example in the
arts, cultural conservation and the like.

DINGO's development was also driven by the goal of being rich enough to

\begin{enumerate}
\def\labelenumi{\arabic{enumi}.}
\tightlist
\item
  integrate and accommodate existing systems and data instances 
\item
  satisfy complex as well as simple use cases, also by straightforward
  extension.
\end{enumerate}

This set of principal design goals and requirements also allowed to work
toward the realisation of additional (and important) objectives, such as
promoting the opening up of funding data, and the linking and re-using
of data.

Special care has been devoted to minimizing the efforts in
applying/adopting the model by users. In particular, while the model has
been created using Linked Data fundamentals, it is apt to different
implementations and integration in non-graph-type data bases, hence it
does not address specifically the optimization of graph inference and
graph-based queries.

\hypertarget{sec:ontologyDesign}{%
\subsection{Approach to Ontology Design}\label{sec:ontologyDesign}}

Ontology generation is a complex process that has been scrutinised in
the literature and has led to the establishment of a number of
engineering best practices, see for example \citep{FLG1997},
\citep{U2000}, \citep{R1999}, \citep{G1997}, \citep{G1998},
\citep{PG2008}. The design of DINGO has followed such best practices.
The main guidelines followed have been:

\begin{itemize}
\item
  a mixture of middle-out and bottom-up approach: starting from actual
  data (such as funding data from various agencies, see below), several
  main concepts have been designed and the ontology generation has
  proceeded by distinguishing a number of commonalities
  (generalisations) and specificities; the advise of domain experts has
  also been essential, mostly profiting from the fact that DINGO has
  been developed at the ERC(EA) \citep{ERC}
\item
  practical usability of the end results
\item
  interoperability/integration from the inception with other graphs (for
  instance, Wikidata and Schema.org)
\item
  sufficient granularity to allow for efficient monitoring and
  evaluation purposes, but also sufficient generality to accommodate
  potentially all funding data, thus providing the whole benefit of a
  large Linked Data Graph. DINGO is straightforwardly extensible to
  provide additional granularity
\item
  coverage of all areas of interest, also for non-academic actors and
  stakeholders.
\end{itemize}

For DINGO's data-based mixed middle-out and bottom-up development we
have used various research funding data, in particular looking at data
freely provided by several funding agencies. For instance, we have used
data from the European Union Funding (Research Framework Programmes),
The Australian Research Council (ARC), the Swiss National Science
Foundation (SNSF), the Croatian Science Foundation, the US National
institute of Health (NIH), the US National Science Foundation (NSF), the
various UK agencies coordinated by the Research Councils UK (RCUK).

Finally, we have adopted elements of agile development, not dissimilarly
from what proposed in \citep{P2017}, for instance concerning unit
testing.

The tools employed in the design and coding process of DINGO have been:
UMLet \citep{UMLET} with some custom diagrams elements for graphical
representations, while the documentation has been build using a custom
software written in Python (unpublished) to automatically generate
human-readable HTML documentation from OWL ontologies serialisations
(see section \ref{sec:documentationSerialisations}).

\hypertarget{sec:ontologyDescription}{%
\subsection{Ontology Description}\label{sec:ontologyDescription}}

Here we describe DINGO's main components and their features, while the
ontology full specification is available at
\href{https://w3id.org/dingo}{\emph{https://w3id.org/dingo}}.

DINGO is an OWL-DL ontology comprising 40 classes and 68 properties. Its
classes provide an articulated conceptualisation of entities relevant
for the characterisation of data in the research, funding and
research-related domain. In particular, besides classes for Projects,
Grants, FundingAgency and others, there are specific classes for
describing funding policies, with several specific subclasses (which can
straightforwardly be expanded).

As we said, the variety and diversity of funding realities (which we
will also call ``realisations'') makes semantic conceptualisation
particularly difficult. For example, different funding agencies/funders
classify their funding policies in various and discording ways,
sometimes using the same word for different things (for instance, the
terms funding scheme/programme/action). Also the role and
characterisation of the different actors in projects and grant
agreements are quite diverse. Such modeling complexity appears not only
at the level of concepts, but also of relations/properties. Notably, the
relationships between the funding and the research enterprise can be
various and rather complex.

Furthermore, alongside concepts definition, additional complexity is
given by the variety of use cases: besides the simple case of one grant
funding one project, often multiple fundings are attached to a single
project (either in sequence or at the same time), or a single grant
funds several (sub)projects.

Therefore, DINGO's properties and classes have been designed to allow
high modeling capability to represent such variety of concepts and
realisations.

DINGO's main features are as follows:

\begin{itemize}
\item
  it defines a number of principal classes: \textbf{Project},
  \textbf{Grant}, \textbf{Funding Agency}, \textbf{FundingScheme},
  \textbf{Role}, \textbf{Person}, \textbf{Organisation},
  \textbf{Criterion}, various subclasses of those and some related
  specialised classes;
\item
  a \textbf{Project} is an organised endeavour (collective or
  individual) planned to reach a particular aim or achieve a result
\item
  a \textbf{Grant} is a disbursed fund paid to a recipient or
  \textbf{beneficiary} and the process for it; DINGO focuses on the main
  definition of ``funding'' (which is defined as ``money for a
  particular purpose; the act of providing money for such a purpose''
  both in the Cambridge, Oxford and Collins dictionaries \citep{COFUND},
  \citep{CAFUND}, \citep{OXFUND}), but can be extended to other types of
  funding (non-monetary ones), see Section
  \ref{sec:maintenanceEvolution}.
\item
  a \textbf{Project} may be funded by one or more Grants simultaneously
  or in sequence
\item
  a \textbf{Grant} may fund one or several Projects
\item
  \textbf{Grants} can be awarded to Person(s) and/or to Organisation(s)
\item
  \textbf{Projects} can be participated by \textbf{Person}(s) or by
  \textbf{Organisation}(s), hence a \textbf{participant}, characterised
  by a \textbf{Role}, can be a \textbf{Person} or an
  \textbf{Organisation}
\item
  the \textbf{Role} class can be used to specify the semantics of the
  participation to a Project or role in a Grant. This class provides
  instruments to model a large variety of semantic types, to account for
  the variety of practices found in actual data
\item
  types of organisations can be specified using one of the several
  sub-classes of \textbf{Organisation} or creating new ones
\item
  a \textbf{participant} (\textbf{Person} or \textbf{Organisation}) in a
  \textbf{Project} may not actually be \textbf{beneficiary} of a
  specific \textbf{Grant} funding the Project; accordingly, DINGO
  reflects that particular participants of Project and beneficiaries of
  Grant funding the same Project may be different
\item
  temporal aspects of the various concepts can be fully modeled, and are
  expressed by specific properties (\emph{start\_time},
  \emph{end\_time}, \emph{inception}, and so on)
\item
  \textbf{Funding Agencies} are the organisations materially disbursing
  and administering the Grant process
\item
  \textbf{Funding Schemes} are funding instruments accompanied by
  specifications of Grant coverage, eligibility, reimbursement rates,
  specific criteria for funding, grant population targets, and similar
  features. Such specifications constitute one or more
  \textbf{Criterion} to award funds (Grants);
\item
  \textbf{Funding Schemes} may be sub-specifications of other Funding
  schemes; this recursive relation allows to model existing complicated
  hierarchies of funding instruments. The word ``Scheme'' has different
  meanings for different funding agencies/funders. In fact, there exist
  other related terms such as funding program and funding action, in
  particular in case of a hierarchy of funding instruments. DINGO
  represents the generalisation of such instruments via the class
  FundingScheme, and expresses the taxonomy and relations among the
  various instruments via the Criterion class and subclasses and the
  FundingScheme (recursive ) class properties
\item
  \textbf{Criteria} can be of different nature, modeled in DINGO via
  different sub-classes; multiple criteria can coexist in a single
  funding scheme; they provide a conceptualisation (straightforwardly
  extensible by sub-classing) to characterise funding policies in
  relation to funding schemes and activities.
\end{itemize}

\begin{figure}
\hypertarget{fig:dingograph}{%
\centering
\includegraphics[width=1.1\textwidth,height=1.3\textheight]{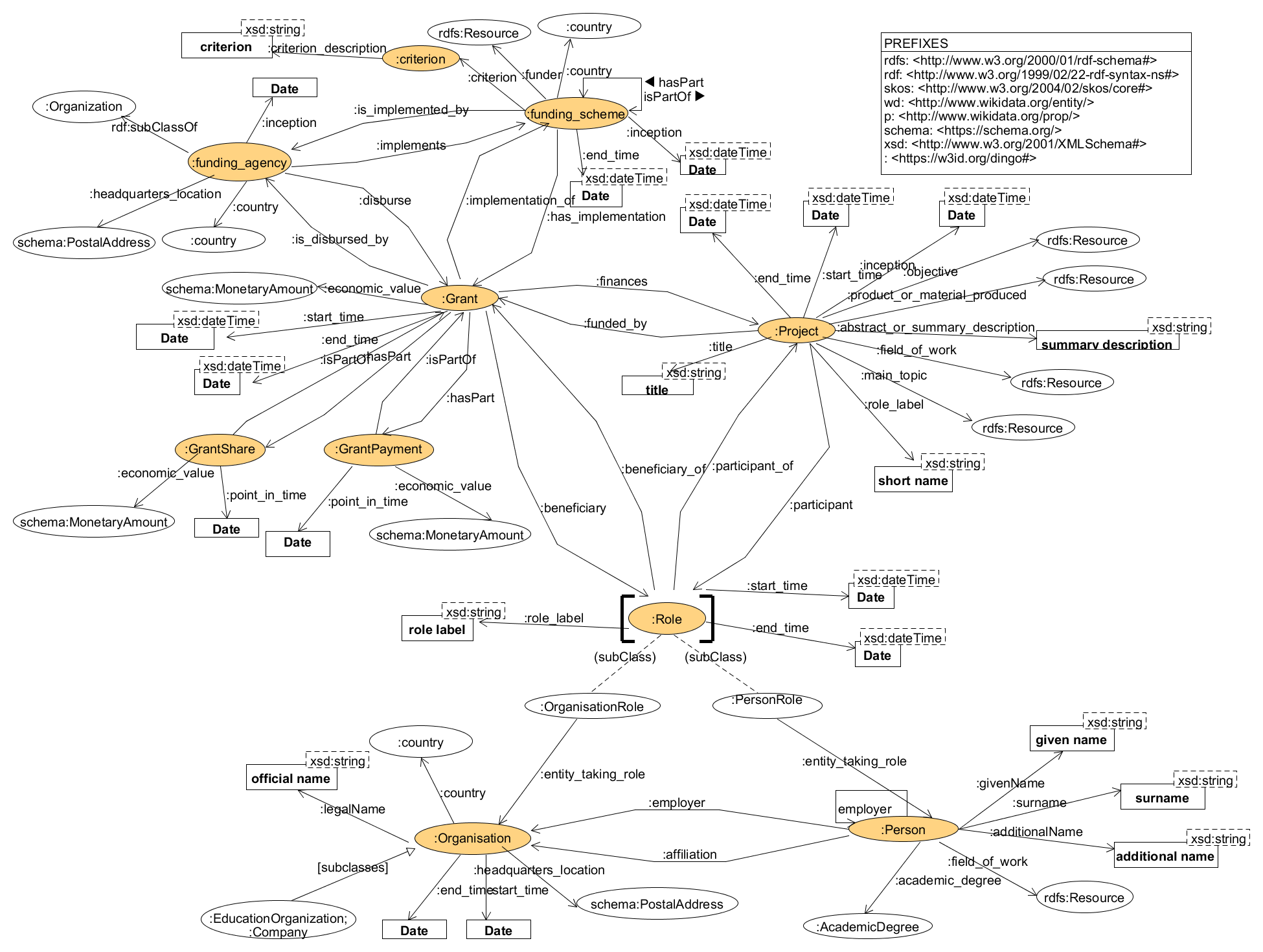}
\caption{Graphical representation of DINGO (main
parts).}\label{fig:dingograph}
}
\end{figure}

We present in Figure \ref{fig:dingograph} a graphical illustration of
the main parts of the ontology, both classes and properties, portrayed
respectively by ovals and arrows.

\hypertarget{sec:documentationSerialisations}{%
\subsection{DINGO Documentation and Machine-readable
Serialisations}\label{sec:documentationSerialisations}}

DINGO is documented at \url{https://w3id.org/dingo}. The documentation
has been created using custom software written in Python (unpublished)
that automatically extracts classes, properties, individuals,
annotations, axioms and namespaces from OWL ontologies and produces
human-readable HTML.

The machine-readable serialisation of DINGO is provided in RDF-Turtle
language, and available at
\href{https://w3id.org/dingo}{\emph{https://w3id.org/dingo}} by
redirection when visited with the ``text/turtle'' header. We also
provide, at the same address, a Shape Expression \citep{SHEX} data model
for validation of data triples.

\hypertarget{sec:maintenanceEvolution}{%
\section{Maintenance and Evolution of
DINGO}\label{sec:maintenanceEvolution}}

DINGO's maintenance is continuous and evolutive in nature, because DINGO
aims at effectively modeling funding and research practices, which
continuously evolve by themselves. As mentioned, the evolution and
extension of DINGO will be eased by the specific design choices made in
creating it, which provide for a high modeling power to cope with the
variety of existing funding realities. Hence, in many cases the required
evolution/extension will be minimal (just by subclassing for new
concepts).

DINGO can however be straightforwardly extended even in more orthogonal
directions. For example, as discussed in Section
\ref{sec:ontologyDescription}, DINGO focuses on the main definition of
``funding'' (the monetary one, see the Cambridge, Oxford and Collins
dictionaries \citep{COFUND}, \citep{CAFUND}, \citep{OXFUND}), but it can
be extended to non-monetary funding simply by providing parallel classes
as Grant, with properties for the specific resources provisions (and
possibly a generalisation class to describe their commonalities).

\hypertarget{sec:conclusion}{%
\section{Summary and Outlook}\label{sec:conclusion}}

We have presented an OWL-based ontology for research and funding called
DINGO and illustrated its main features, uptake and evolutive
maintenance.

DINGO has the potential to constitute a key ingredient for a set of
orthogonal and interoperable ontologies for the knowledge area of
funding, research and their impact. In particular, there is a lack of
ontological conceptualisations concerning the domain of impact and
impact studies, hence, for instance, we have already planned the
development of ontologies for data relative to impact indicators.

Moreover, as we mentioned, DINGO has features that enable it to be both
used for domain knowledge graphs specific to research, as well as in
graphs for other domains where funding aspects and policies are of
interest (such as the arts, cultural conservation, and the like).

DINGO has already been used in a number of projects, as described in
Section \ref{sec:communityUptake}. We plan to engage further with
relevant communities to create systems that offer information on
research funding in distributed manner using DINGO. This should
eventually lead to a truly global Open Research Information Graph
providing access to data in several interconnected research information
systems.

\textbf{Acknowledgements.} We would like to thank the co-organisers and
the participants of the workshop ``Wikidata for research'', Berlin,
17-18 June 2018 for their feedback and input.

\bibliography{DINGOontologyPaper}

\end{document}